\documentclass[aps,pra,twocolumn,twoside,a4paper,10pt,amsmath,amssymb,showpacs]{revtex4-1}

\usepackage{mathrsfs}
\usepackage{graphicx}
\usepackage{CJK}

\def\bra#1{\left\langle{#1}\right|}
\def\ket#1{\left|{#1}\right\rangle}
\def\braket#1#2{\left\langle{{#1}}\mathrel{\left|{\vphantom{{#1}{#2}}}\right.\kern-\nulldelimiterspace}{{#2}}\right\rangle}

\def\romanNum#1{\romannumeral#1}
\def\RomanNum#1{\textrm{\uppercase\expandafter{\romannumeral#1}}}

\begin{document}

\title{Clock frequency estimation under spontaneous emission}

\author{Xizhou Qin$^{1}$}
\author{Jiahao Huang$^{1}$}
\author{Honghua Zhong$^{1}$}
\author{Chaohong Lee$^{1,2,}$}
\altaffiliation{Corresponding author. Email: lichaoh2@mail.sysu.edu.cn}

\affiliation{$^{1}$TianQin Research Center \& School of Physics and Astronomy, Sun Yat-Sen University (Zhuhai Campus), Zhuhai 519082, China}
\affiliation{$^{2}$Key Laboratory of Optoelectronic Materials and Technologies, Sun Yat-Sen University (Guangzhou Campus), Guangzhou 510275, China}

\date{\today}

\begin{abstract}
We investigate the quantum dynamics of a driven two-level system under spontaneous emission and its application in clock frequency estimation.
By using a Lindblad equation to describe the system, we analytically obtain its exact solutions, which show three different regimes: Rabi oscillation, damped oscillation and overdamped decay.
From the analytical solutions, we explore how the spontaneous emission affects the clock frequency estimation.
We find that, under a modest spontaneous emission rate, the transition frequency can still be inferred from the Rabi oscillation.
Our results provide potential practical applications in frequency measurement and quantum control under decoherence.
\\ \\
\textbf{Keywords:} clock frequency estimation, two-level system, spontaneous emission
\pacs{03.65.Yz, 42.50.Ct, 06.20.-f}

\end{abstract}

\maketitle

\section{Introduction\label{Sec1}}

Clock frequency estimation forms the basis of physics for precision measurements~\cite{Hall2006,Hansch2006,Margolis2014}, spectroscopy~\cite{Takamoto2003}, and astronomy~\cite{Steinmetz2008} etc.
It is also the heart of the modern technologies, from global position system (GPS)~\cite{Grewal2013} and magnetometers~\cite{Kitching2011,Budker2007} to inertial sensors~\cite{Kitching2011} (e.g., gyroscopes and gravimeters).
A two-level system (TLS) coupled with an oscillating field is the prototype of a clock.
This may correspond to an ensemble of atoms or an ion driven by an electromagnetic field in the microwave, optical or ultraviolet regions.
The TLS's tick with the oscillations of the electromagnetic wave whose frequency is locked to that of the energy difference between two discrete levels.
The closer the frequency of the external field to the transition frequency of the TLS, the larger contrast of the Rabi oscillation occurs.
Hence, the clock frequency can be estimated according to the TLS's population dynamics.
This model was first proposed by Rabi~\cite{Rabi1937} and widely used in atomic clocks~\cite{Hinkley2013,Bloom2014,Nicholson2015}.

The performances of an atomic clock are affected by many factors~\cite{Porsev2006,Taichenachev2006,Gibble2009}.
In realistic experiments, decoherence is unavoidable and plays an important role, especially when the uncertainty is at the level of $10^{-17}$~\cite{Rosenband2008,Chou2010} or even smaller~\cite{Huntemann2016}.
For an atomic clock, one of the most essential mechanisms of decoherence is spontaneous emission.
The effects of spontaneous emission may have great influence on the TLS's population dynamics and eventually degrade the performance of the clock.
Therefore, it is of great interest to investigate how the spontaneous emission affects the population dynamics and the clock frequency estimation.

Theoretically, a TLS driven by an oscillating field under spontaneous emission is well described by a Lindblad equation~\cite{Weinberg2016}.
Although the driven TLS's are extensively investigated, the search for exactly solvable driven TLS's is still an active field~\cite{DasSarma2012}.
Two famous examples of solvable driven TLS's are the Landau-Zener~\cite{Landau1932,Zener1932} and Rabi~\cite{Rabi1937} problems.
Solvable models on driven TLS's provide the reliable explanations for experimental observations as well as numerical calculations in quantum optics~\cite{AllenBook1992,BoydBook1992,MeystreBook2001,HarocheBook2006}, quantum metrology~\cite{Ivanov2013,Malossi2013,Tian2015,Ivanov2015}, and quantum control~\cite{Greilich2009,Poem2011,Bason2012,Sauer2013,Daems2013}.
Among the existed analytical methods for solving TLS's~\cite{DasSarma2012,Wu2007,Wu20072,Xie2010,Hai2013,Barnes2013,Messina2014}, much attention is paid to the driven TLS's described by Schr\"{o}dinger equations which fail to include the spontaneous emission.
The solvable driven TLS obeying Lindblad equation~\cite{BreuerBook2002,NielsenBook2000} is still of rareness (see Ref.~\cite{HanggiRev1998} for a review on dissipative driven TLS's).
Furthermore, the analytical study of decoherence is always hard and nontrivial for practical applications~\cite{Thorwart2000,Mogilevtsev2008,Sergi2013,Sauer2013,Zlatanov2015}.

Here, we study a TLS driven by an oscillating field under spontaneous emission.
Within the Markovian approximation, the system can be described by a Lindblad master equation.
Then, we solve the Lindblad equation analytically and give its exact solutions.
We analyze different regimes for the population dynamics and investigate the clock frequency estimation in the presence of spontaneous emission.
Our analytical results illustrate that, the spontaneous emission does not induce any frequency shift but the estimated precision of the transition frequency would be reduced.

This article is organized as follows.
In Sec.~\ref{Sec2}, we give the Bloch equations for our system.
In Sec.~\ref{Sec2A}, by implementing a linear transformation, we obtain an ordinary differential equations with constant coefficients for the system.
In Sec.~\ref{Sec2B} and~\ref{Sec2C}, we give the analytical solutions for the Lindblad equation.
In Sec.~\ref{Sec3}, we discuss the population dynamics and the clock frequency estimation in the presence of spontaneous emission.
Finally, a brief summary of our results is given in Sec.~\ref{Sec4}.

\section{Lindblad equation and its analytical solutions\label{Sec2}}

\begin{figure}[t]
  \includegraphics[width=1.0\columnwidth]{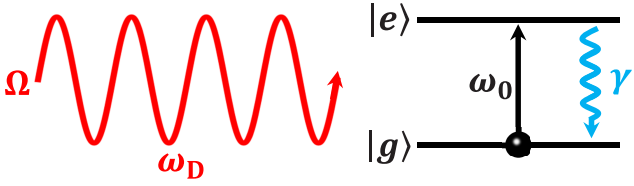}
  \caption{\label{Fig1}(Color online). Schematic diagram of a TLS coupled with an oscillating electromagnetic field under spontaneous emission. Here, $\Omega$ and $\omega_\mathrm{D}$ are the Rabi frequency and oscillation frequency of the driving field, $\omega_0$ is the transition frequency of the TLS and $\gamma$ stands for the spontaneous emission rate.}
\end{figure}

We consider two stable states $\ket{g}$ and $\ket{e}$ in an atomic clock.
By introducing the pseudospin-$1/2$ operators
$\hat S^+=\ket{e}\bra{g}$,
$\hat S^-=\ket{g}\bra{e}$,
$\hat S^x=\frac{1}{2}(\hat S^++\hat S^-)$,
$\hat S^y=\frac{1}{2i}(\hat S^+-\hat S^-)$,
and $\hat S^z=\frac{1}{2}(\ket{e}\bra{e}-\ket{g}\bra{g})$,
within the Markovian approximations, the density matrix operator of the system, $\hat\rho(t)$, is governed by a Lindblad equation (throughout this article, we set $\hbar=1$),
\begin{equation}\label{Eq_lindblad_S}
  \tfrac{\mathrm d}{\mathrm dt}\hat\rho(t)=-i\big[\hat H(t),\hat\rho(t)\big]
  +\mathcal{L}\big[\hat\rho(t)\big].
\end{equation}
The time-dependent Hamiltonian reads,
\begin{eqnarray}
  \hat H(t)&=&\hat H_0+\hat H_1(t), \\
  \hat H_0&=&\omega_0\hat S^z, \\
  \hat H_1(t)&=&\Omega\big[\hat S^x\cos(\omega_\mathrm{D}t)+\hat S^y\sin(\omega_\mathrm{D}t)\big],
\end{eqnarray}
where $\hat H_0$ is the Hamiltonian of the TSL with transition frequency $\omega_0$,
and $\hat H_1(t)$ corresponds to the coupling between the TSL and the external electromagnetic field with Rabi frequency $\Omega$ and driving frequency $\omega_\mathrm{D}$.
The schematic is shown in Fig.~\ref{Fig1}.
Here, $\mathcal{L}$ is a superoperator which describes the spontaneous emission~\cite{BreuerBook2002,NielsenBook2000},
\begin{equation}\label{Eq_SO_S}
  \mathcal{L}\big[\hat\rho\big]=\gamma\big(\hat S^-\hat\rho\hat S^+
  -\tfrac{1}{2}\hat S^+\hat S^-\hat\rho-\tfrac{1}{2}\hat\rho\hat S^+\hat S^-\big),
\end{equation}
with $\gamma$ being the spontaneous emission rate.

In this article, we derive the equations in the Dirac picture.
By applying the transformation $\hat\rho^\mathrm{D}(t)=e^{i\hat H_0t}\hat\rho(t)e^{-i\hat H_0t}$,
the Lindblad equation~\eqref{Eq_lindblad_S} becomes
\begin{equation}\label{Eq_lindblad_D}
  \tfrac{\mathrm d}{\mathrm dt}\hat\rho^\mathrm{D}(t)
  =-i\big[\hat H_1^\mathrm{D}(t),\hat\rho^\mathrm{D}(t)\big]
  +\mathcal{L}^\mathrm{D}\!\big[\hat\rho^\mathrm{D}(t)\big].
\end{equation}
Here, by using the identity $e^{i\hat H_0 t}\hat S^\pm e^{-i\hat H_0 t}=e^{\pm i\omega_0 t}\hat S^\pm$,
\begin{equation}\label{Eq_Driving_D}
  \hat H_1^\mathrm{D}(t)=\tfrac{1}{2}\Omega\big(\hat S^+e^{-i\Delta t}+\hat S^-e^{i\Delta t}\big)
\end{equation}
and
\begin{equation}\label{Eq_SO_D}
  \mathcal{L}^\mathrm{D}\!\!\left[\hat\rho^\mathrm{D}\right]=\gamma\big(
  \hat S^-\hat\rho^\mathrm{D}\hat S^+-\tfrac{1}{2}\hat S^+\hat S^-\hat\rho^\mathrm{D}
  -\tfrac{1}{2}\hat\rho^\mathrm{D}\hat S^+\hat S^-\big),
\end{equation}
where the detuning is defined as $\Delta=\omega_\mathrm{D}-\omega_0$.
Eq.~\eqref{Eq_lindblad_D} in the basis of eigenstates of $\hat S^z$ becomes
\begin{equation}\label{Eq_Matrix_Form}
  \!\left\{\!\!\begin{array}{l}
  \frac{\mathrm d}{\mathrm dt}\rho^{\mathrm{D}00}
    =\frac{1}{2i}\Omega\left(e^{-i\Delta t}\rho^{\mathrm{D}10}-e^{i\Delta t}\rho^{\mathrm{D}01}\right)-\gamma\rho^{\mathrm{D}00}, \\
  \frac{\mathrm d}{\mathrm dt}\rho^{\mathrm{D}01}
    =\frac{1}{2i}\Omega e^{-i\Delta t}\left(\rho^{\mathrm{D}11}-\rho^{\mathrm{D}00}\right)-\frac{1}{2}\gamma\rho^{\mathrm{D}01}, \\
  \frac{\mathrm d}{\mathrm dt}\rho^{\mathrm{D}10}
    =-\frac{1}{2i}\Omega e^{i\Delta t}\left(\rho^{\mathrm{D}11}-\rho^{\mathrm{D}00}\right)-\frac{1}{2}\gamma\rho^{\mathrm{D}10}, \\
  \frac{\mathrm d}{\mathrm dt}\rho^{\mathrm{D}11}
    =-\frac{1}{2i}\Omega\left(e^{-i\Delta t}\rho^{\mathrm{D}10}-e^{i\Delta t}\rho^{\mathrm{D}01}\right)+\gamma\rho^{\mathrm{D}00},
  \end{array}\right.\!\!\!\!\!\!\!\!
\end{equation}
with
\begin{equation}
  \rho^\mathrm{D}(t)=\left(\begin{array}{cc}
    \rho^\mathrm{D00}(t) & \rho^\mathrm{D01}(t) \\
    \rho^\mathrm{D10}(t) & \rho^\mathrm{D11}(t)
  \end{array}\right).
\end{equation}

Next, introducing the Bloch vector,
$\mathbf{R}=(u,v,w)^\mathrm{T}$ (where the superscript ``T" denotes the operation of matrix transpose),
the density matrix can be written as
\begin{equation}
  \hat\rho^\mathrm{D}=\tfrac{1}{2}\left(\hat 1+u\hat\sigma^x+v\hat\sigma^y+w\hat\sigma^z\right),
\end{equation}
where $\hat\sigma^x$, $\hat\sigma^y$ and $\hat\sigma^z$ are the Pauli matrices.
The three coefficients can be expressed in the form of
\begin{equation}\label{Eq_Pauli_uvw}
  \left\{\!\!\begin{array}{ccccl}
    u&\!\!=\!\!&\mathrm{tr}[\hat\rho^\mathrm{D}\hat\sigma^x]&\!\!=\!\!&
    \cos(\omega_0t)\left<\hat\sigma^x\right>+\sin(\omega_0t)\left<\hat\sigma^y\right>,\\
    v&\!\!=\!\!&\mathrm{tr}[\hat\rho^\mathrm{D}\hat\sigma^y]&\!\!=\!\!&
    -\sin(\omega_0t)\left<\hat\sigma^x\right>+\cos(\omega_0t)\left<\hat\sigma^y\right>,\\
    w&\!\!=\!\!&\mathrm{tr}[\hat\rho^\mathrm{D}\hat\sigma^z]&\!\!=\!\!&
    \left<\hat\sigma^z\right>.
  \end{array}\right.\!\!\!\!\!\!
\end{equation}
Here, $\left<\hat\sigma^\alpha\right>=\mathrm{tr}[\hat\rho(t)\hat\sigma^\alpha]$ is the expectation value of $\hat\sigma^\alpha$ for the evolved state $\hat\rho(t)$ ($\alpha=x,y,z$).
Then, we obtain the Bloch equations
\begin{equation}\label{Eq_Bloch_Original}
  \left\{\!\!\begin{array}{l}
    \frac{\mathrm{d}}{\mathrm{d}t}u
      =-\frac{1}{2}\gamma u+\Omega\sin(\Delta t)w, \\
    \frac{\mathrm{d}}{\mathrm{d}t}v
      =-\frac{1}{2}\gamma v-\Omega\cos(\Delta t)w, \\
    \frac{\mathrm{d}}{\mathrm{d}t}w
      =-\Omega\sin(\Delta t)u+\Omega\cos(\Delta t)v-\gamma w-\gamma.
  \end{array}\right.
\end{equation}
The Bloch equations~\eqref{Eq_Bloch_Original} can be written in a more compact form
\begin{equation}\label{Eq_Bloch_Original_Compact}
  \tfrac{\mathrm{d}}{\mathrm{d}t}\mathbf{R}=\mathbf{M}\mathbf{R}+\mathbf{R}_0,
\end{equation}
where $\mathbf{R}_0=-\gamma(0,0,1)^\mathrm{T}$, and the coefficient matrix is defined as
\begin{equation}\label{Eq_Bloch_M}
  \mathbf{M}=\left(\begin{array}{ccc}
  -\frac{1}{2}\gamma & 0 & \Omega\sin(\Delta t) \\
  0 & -\frac{1}{2}\gamma & -\Omega\cos(\Delta t) \\
  -\Omega\sin(\Delta t) & \Omega\cos(\Delta t) & -\gamma
  \end{array}\right).
\end{equation}

\subsection{Linear transformation\label{Sec2A}}

In general, when the coefficient matrix $\mathbf{M}$~\eqref{Eq_Bloch_M} is time-dependent, it is not easy to solve the Bloch equation~\eqref{Eq_Bloch_Original_Compact} directly.
Such differential equations with periodically oscillating coefficients can be dealt by Floquet theory~\cite{EasthamBook1973}.
However, the obtained solutions are usually given in a form of a series, which may be sometimes inconvenient.
Here, we present another simpler approach, which enables us to obtain the completely and analytical solutions in closed-form expressions.

First, we take a general reversible linear transformation on the Bloch vectors,
\begin{equation}\label{Eq_linear_trans}
  \mathbf{R}=\mathbf{P}\mathbf{R}^Q,
\end{equation}
where $\mathbf{R}^Q=(u^Q,v^Q,w^Q)^\mathrm{T}$ is a new set of variables,
and $\mathbf{P}$ is an undetermined $3 \times 3$ matrix with time-dependent matrix elements.
Meanwhile, Eq.~\eqref{Eq_Bloch_Original_Compact} becomes
\begin{equation}\label{Eq_Bloch_RQ}
  \tfrac{\mathrm{d}}{\mathrm{d}t}\mathbf{R}^Q=\mathbf{F}\mathbf{R}^Q+\mathbf{G},
\end{equation}
with
\begin{eqnarray}\label{FG}
  && \mathbf{F}=\mathbf{P}^{-1}\mathbf{M}\mathbf{P}
  -\mathbf{P}^{-1}\tfrac{\mathrm{d}}{\mathrm{d}t}\mathbf{P}, \\
  && \mathbf{G}=\mathbf{P}^{-1}\mathbf{R}_0.
\end{eqnarray}
The differential equation~\eqref{Eq_Bloch_RQ} can be simplified into a linear one with constant coefficients
if $\mathbf{F}$ is time-independent,
or equivalently $\tfrac{\mathrm{d}}{\mathrm{d}t}\mathbf{F}=0$.
To let $\tfrac{\mathrm{d}}{\mathrm{d}t}\mathbf{F}=0$, it is not unique to choose the transformation matrix $\mathbf{P}$.
Here, as an example, we set $\mathbf{P}$ in the form of
\begin{equation}\label{Eq_linear_trans_P}
  \mathbf{P}=e^{-\gamma t}
  \left(\begin{array}{ccc}
  \cos(\Delta t) & -\sin(\Delta t) & 0 \\
  \sin(\Delta t) & \cos(\Delta t) & 0 \\
  0 & 0 & 1
  \end{array}\right),
\end{equation}
according to $\mathbf{M}$~\eqref{Eq_Bloch_M}.
Substituting $\mathbf{P}$~\eqref{Eq_linear_trans_P} into Eq.~\eqref{FG}, one can easily obtain
\begin{equation}\label{Eq_Matrix_F}
  \mathbf{F}=\left(\begin{array}{ccc}
  \frac{1}{2}\gamma & \Delta & 0 \\
  -\Delta & \frac{1}{2}\gamma & -\Omega \\
  0 & \Omega & 0
  \end{array}\right)
\end{equation}
and $\mathbf{G}=-\gamma e^{\gamma t}(0,0,1)^\mathrm{T}$, respectively. Here, $\mathbf{F}$~\eqref{Eq_Matrix_F} satisfies the desired condition $\tfrac{\mathrm{d}}{\mathrm{d}t}\mathbf{F}=0$.
Until by now, we have simplified the Bloch equation~\eqref{Eq_Bloch_Original_Compact} by a straight linear transformation and the new differential equation can be solved much more easily.

\subsection{Solutions with zero detuning\label{Sec2B}}

At resonance $\omega_D=\omega_0$ (i.e., zero detuning $\Delta=0$), Eq.~\eqref{Eq_Bloch_RQ} can be reduced to
\begin{equation}\label{Eq_Bloch_Zero_detuning}
  \left\{\!\!\begin{array}{l}
    \frac{\mathrm{d}}{\mathrm{d}t}u^Q=\frac{1}{2}\gamma u^Q, \\
    \frac{\mathrm{d}}{\mathrm{d}t}v^Q=\frac{1}{2}\gamma v^Q-\Omega w^Q, \\
    \frac{\mathrm{d}}{\mathrm{d}t}w^Q=\Omega v^Q-\gamma e^{\gamma t}.
  \end{array}\right.
\end{equation}
In this case, it is easy to find that $u^Q(t)=u_0 e^{\frac{1}{2}\gamma t}$, $v^Q(t)=\frac{1}{\Omega}[\frac{\mathrm d}{\mathrm dt}w^Q(t)+\gamma e^{\gamma t}]$,
and
\begin{equation}\label{Eq_Zero_detuning_f}
  \tfrac{\mathrm d^2}{\mathrm dt^2}w^Q(t)
  -\tfrac{1}{2}\gamma\tfrac{\mathrm d}{\mathrm dt}w^Q(t)
  +\Omega^2 w^Q(t)
  +\tfrac{1}{2}\gamma^2 e^{\gamma t}=0.
\end{equation}
A special solution to Eq.~\eqref{Eq_Zero_detuning_f} is
$w^Q_s(t)=-\tfrac{\gamma^2}{\gamma^2+2\Omega^2} e^{\gamma t}$.
The characteristic equation of Eq.~\eqref{Eq_Zero_detuning_f} is
\begin{equation}\label{Eq_Zero_detuning_chara}
  \lambda^2-\tfrac{1}{2}\gamma\lambda+\Omega^2=0,
\end{equation}
We denote the discriminant of~\eqref{Eq_Zero_detuning_chara} as
$D_0=\frac{1}{4}\gamma^2-4\Omega^2$,
then, the general solution of Eq.~\eqref{Eq_Zero_detuning_f} is \\
(\romanNum{1}) for $D_0>0$ ($\gamma>4\Omega$),
\begin{equation}\label{Eq_Bloch_Zero_D0>0}
  w^Q(t)=C_1e^{\lambda_1t}+C_2e^{\lambda_2t}+w^Q_s(t);
\end{equation}
(\romanNum{2}) for $D_0=0$ ($\gamma=4\Omega$),
\begin{equation}\label{Eq_Bloch_Zero_D0=0}
  w^Q(t)=e^{\eta t}(C_1+C_2t)+w^Q_s(t);
\end{equation}
(\romanNum{3}) for $D_0<0$ ($0\le\gamma<4\Omega$),
\begin{equation}\label{Eq_Bloch_Zero_D0<0}
  w^Q(t)=e^{\eta t}\big[C_1\cos(\omega t)+C_2\sin(\omega t)\big]+w^Q_s(t).
\end{equation}
Here, in Eq.~\eqref{Eq_Bloch_Zero_D0>0},
$\lambda_{1,2}=\frac{1}{4}\gamma\pm\frac{1}{2}\sqrt{D_0}$ ($\lambda_1>\lambda_2$)
are two distinct roots of Eq.~\eqref{Eq_Zero_detuning_chara};
in Eq.~\eqref{Eq_Bloch_Zero_D0=0}, $\eta=\frac{1}{4}\gamma$ is the double root;
and in Eq.~\eqref{Eq_Bloch_Zero_D0<0}, $\eta\pm i\omega=\frac{1}{4}\gamma\pm i\frac{1}{2}\sqrt{-D_0}$
are two conjugate complex roots
($\eta=\frac{1}{4}\gamma$ and $\omega=\frac{1}{2}\sqrt{-D_0}$).
The coefficients ($C_1$,~$C_2$) appear in Eqs.~\eqref{Eq_Bloch_Zero_D0>0}-\eqref{Eq_Bloch_Zero_D0<0}
are determined via the initial conditions (see Appendix~\ref{SecApp1}).
So far, we completely solve the Bloch equation~\eqref{Eq_Bloch_Original_Compact} in the zero detuning case ($\Delta=0$).

\subsection{Solutions with nonzero detuning\label{Sec2C}}

For the case of nonzero detuning ($\Delta\ne 0$), Eq.~\eqref{Eq_Bloch_RQ} becomes
\begin{equation}\label{Eq_Bloch_NoneZero_detuning}
  \left\{\!\!\begin{array}{l}
    \frac{\mathrm{d}}{\mathrm{d}t}u^Q=\frac{1}{2}\gamma u^Q+\Delta v^Q, \\
    \frac{\mathrm{d}}{\mathrm{d}t}v^Q=-\Delta u^Q+\frac{1}{2}\gamma v^Q-\Omega w^Q, \\
    \frac{\mathrm{d}}{\mathrm{d}t}w^Q=\Omega v^Q-\gamma e^{\gamma t}.
  \end{array}\right.
\end{equation}
%
%
Here, we have $v^Q(t)=\frac{1}{\Omega}[\frac{\mathrm d}{\mathrm dt}w^Q(t)+\gamma e^{\gamma t}]$,
$u^Q(t)=\frac{1}{\Delta\Omega}[-\frac{\mathrm{d}^2}{\mathrm{d}t^2}w^Q(t)
+\frac{1}{2}\gamma\frac{\mathrm d}{\mathrm dt}w^Q(t)-\Omega^2w^Q(t)
-\frac{1}{2}\gamma^2e^{\gamma t}]$, and
\begin{eqnarray}\label{Eq_Nonzero_detuning_f}
  \tfrac{\mathrm{d}^3}{\mathrm{d}t^3}w^Q(t)
    &-&\gamma\tfrac{\mathrm{d}^2}{\mathrm{d}t^2}w^Q(t)
    +\left(\Delta^2+\Omega^2+\tfrac{1}{4}\gamma^2\right)\tfrac{\mathrm{d}}{\mathrm{d}t}w^Q(t)
    \nonumber \\
    &-&\tfrac{1}{2}\Omega^2\gamma w^Q(t)
    +\gamma\left(\Delta^2+\tfrac{1}{4}\gamma^2\right)e^{\gamma t}=0.
\end{eqnarray}
A special solution to Eq.~\eqref{Eq_Nonzero_detuning_f} is
$w^Q_s(t)=-\tfrac{4\Delta^2+\gamma^2}{4\Delta^2+\gamma^2+2\Omega^2}e^{\gamma t}$.
The characteristic equation of Eq.~\eqref{Eq_Nonzero_detuning_f} is
\begin{equation}\label{Eq_Nonzero_detuning_chara}
  \lambda^3-\gamma\lambda^2+\left(\Delta^2+\Omega^2+\tfrac{1}{4}\gamma^2\right)\lambda-\tfrac{1}{2}\Omega^2\gamma=0.
\end{equation}
Let $\lambda=\lambda'+\frac{1}{3}\gamma$, the characteristic equation~\eqref{Eq_Nonzero_detuning_chara} becomes
\begin{equation}\label{Eq_Nonzero_detuning_chara_2}
  \lambda'^3+p\lambda'+q=0,
\end{equation}
with $p=\Delta^2+\Omega^2-\frac{1}{12}\gamma^2$ and $q=\frac{1}{108}\gamma(36\Delta^2-18\Omega^2+\gamma^2)$.
We denote the discriminant of \eqref{Eq_Nonzero_detuning_chara_2} as $D=\frac{1}{4}q^2+\frac{1}{27}p^3
=\frac{1}{432}[\Delta^2\gamma^4+(8\Delta^4-20\Delta^2\Omega^2-\Omega^4)\gamma^2+16(\Delta^2+\Omega^2)^3]$, then the general solution of Eq.~\eqref{Eq_Nonzero_detuning_f} is \\
(\romanNum{1}) for $D=0$ and $p=0$,
\begin{equation}\label{Eq_Bloch_Nonzero_1}
  w^Q(t)=e^{\lambda_1 t}(C_1+C_2t+C_3t^2)+w^Q_s(t);
\end{equation}
(\romanNum{2}) for $D=0$ and $p\ne0$,
\begin{equation}\label{Eq_Bloch_Nonzero_2}
  w^Q(t)=C_1e^{\lambda_1 t}+e^{\lambda_2 t}(C_2+C_3t)+w^Q_s(t);
\end{equation}
(\romanNum{3}) for $D<0$,
\begin{equation}\label{Eq_Bloch_Nonzero_3}
  w^Q(t)=C_1e^{\lambda_1 t}+C_2e^{\lambda_2 t}+C_3e^{\lambda_3 t}+w^Q_s(t);
\end{equation}
(\romanNum{4}) for $D>0$,
\begin{equation}\label{Eq_Bloch_Nonzero_4}
  w^Q(t)=C_1e^{\lambda_1 t}+e^{\eta t}\big[C_2\cos(\omega t)+C_3\sin(\omega t)\big]+w^Q_s(t).
\end{equation}
Here, in Eq.~\eqref{Eq_Bloch_Nonzero_1}, $\lambda_1=\frac{1}{3}\gamma$ is the triple root of Eq.~\eqref{Eq_Nonzero_detuning_chara};
in Eq.~\eqref{Eq_Bloch_Nonzero_2}, $\lambda_1=\frac{1}{3}\gamma-(4q)^{1/3}$ is the single root and $\lambda_2=\frac{1}{3}\gamma+(\frac{1}{2}q)^{1/3}$ the double one;
in Eq.~\eqref{Eq_Bloch_Nonzero_3}, $\lambda_1=\frac{1}{3}\gamma+r\cos(\frac{1}{3}\phi)$, $\lambda_2=\frac{1}{3}\gamma+r\cos[\frac{1}{3}(\phi+2\pi)]$, $\lambda_3=\frac{1}{3}\gamma+r\cos[\frac{1}{3}(\phi-2\pi)]$ are the three distinct real roots [where, $r=2(-\frac{1}{3}p)^{1/2}$ and $\phi=\arccos(-4qr^{-3})$];
in Eq.~\eqref{Eq_Bloch_Nonzero_4}, $\lambda_1=\frac{1}{3}\gamma+R_1+R_2$ is the real root while $\eta\pm i\omega$ the two conjugate complex roots with $\eta=\frac{1}{3}\gamma-\frac{1}{2}(R_1+R_2)$ and $\omega=\frac{\sqrt{3}}{2}(R_1-R_2)$ [where $R_1=(-\frac{1}{2}q+\sqrt{D})^{1/3}$, $R_2=(-\frac{1}{2}q-\sqrt{D})^{1/3}$].
The coefficients ($C_1$,~$C_2$,~$C_3$) appear in Eqs.~\eqref{Eq_Bloch_Nonzero_1}-\eqref{Eq_Bloch_Nonzero_4}
are determined via the initial conditions (see Appendix~\ref{SecApp2}).
Till here, we completely solve the Bloch equation~\eqref{Eq_Bloch_Original_Compact} in the nonzero detuning case ($\Delta\neq0$).

\section{Clock frequency estimation\label{Sec3}}

In the previous section, we have mathematically solved the Bloch equation~\eqref{Eq_Bloch_Original_Compact} for both zero ($\Delta=0$) and nonzero ($\Delta\ne 0$) detuning cases and found out their exact analytical solutions.
In this section, we mainly discuss the physics behind the analytical solutions and show how the spontaneous emission affects the clock frequency estimation within our model.

\subsection{Rabi oscillating, Damped oscillating and overdamping regimes\label{Sec3A}}

In the absence of spontaneous emission ($\gamma=0$), the TLS performs the Rabi oscillation.
The population difference $w(t)$ evolves as
\begin{eqnarray}\label{Eq_gamma_0}
  w(t)&=&\tfrac{1}{\omega_\mathrm{R}^2}\Big[
  \Delta(-\Omega u_0+\Delta w_0)
  +\omega_\mathrm{R}\Omega v_0\sin(\omega_\mathrm{R} t)\nonumber \\
  &&+\Omega(\Delta u_0+\Omega w_0)\cos(\omega_\mathrm{R} t)\Big],
\end{eqnarray}
with the total Rabi frequency $\omega_\mathrm{R}=\sqrt{\Delta^2+\Omega^2}$.
For the initial state $\rho(0)=\ket{g}\bra{g}$, one has $\rho^{\mathrm{D}}(0)=\rho(0)$,
and $(u_0,v_0,w_0)^{\mathrm T}=(0,0,-1)^{\mathrm T}$.
The solution for $w(t)$~\eqref{Eq_gamma_0} is reduced to
\begin{equation}
  w(t)=-\frac{\Delta^2}{\omega_\mathrm{R}^2}
  -\frac{\Omega^2}{\omega_\mathrm{R}^2}\cos(\omega_\mathrm{R} t).
\end{equation}
The excited population in $\ket{e}$ is given as
\begin{equation}
  P_{e}(t)=\frac{1+w(t)}{2}=\frac{\Omega^2}{\omega_\mathrm{R}^2}\sin^2\left(\frac{\omega_\mathrm{R} t}{2}\right),
\end{equation}
which is consistent with the well-known Rabi oscillation.

\begin{figure}[t]
  \includegraphics[width=1.0\columnwidth]{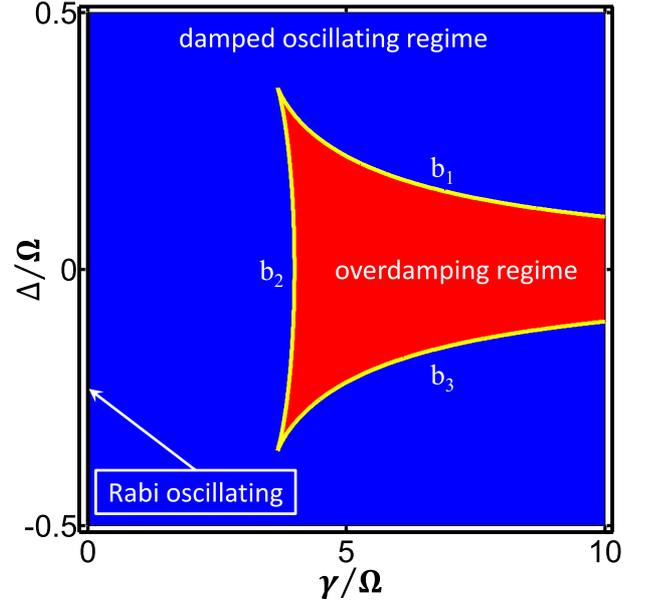}
  \caption{\label{Fig2}(Color online). The phase diagram of three different types of solutions. The black line lying at $\gamma/\Omega=0$ represents the Rabi oscillating solutions. The blue area is the damped oscillating regime. The red area as well as the yellow curves $b_1$, $b_2$, $b_3$, denotes the overdamping regime.}
\end{figure}

\begin{figure*}[t]
  \includegraphics[width=2.0\columnwidth]{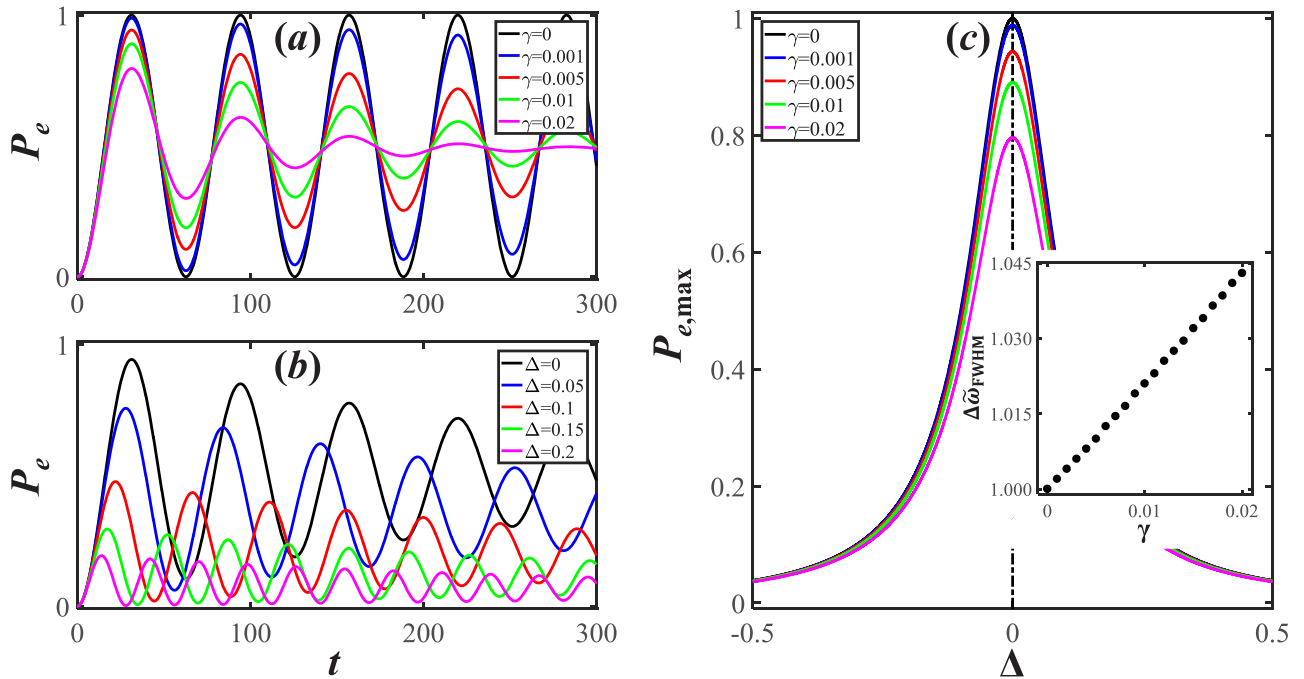}
  \caption{\label{Fig3}(Color online). (a) Evolution of excited population at resonance ($\Delta=0$) under different spontaneous emission rates $\gamma=0$ (black), $0.001$ (blue), $0.005$ (red), $0.01$ (green), and $0.02$ (magenta). (b) Evolution of excited population under spontaneous emission ($\gamma=0.005$) with different detunings $\Delta=0$ (black), $0.05$ (blue), $0.1$ (red), $0.15$ (green), and $0.2$ (magenta). (c) The spectrum of maximal excited population versus detuning $\Delta$ under different spontaneous emission rates $\gamma=0$ (black), $0.001$ (blue), $0.005$ (red), $0.01$ (green), and $0.02$ (magenta). Inset shows the relationship of the relative full width at half maximum (FWHM) of the spectrum $\Delta\tilde\omega_{\text{FWHM}}$ versus spontaneous emission rate $\gamma$. Here, we set $\omega_0=10$ and $\Omega=0.1$.}
\end{figure*}

In the presence of spontaneous emission ($\gamma\ne0$), there are two types of solutions corresponding to the damped oscillating and the overdamping regimes, respectively.
In damped oscillating regime, the excited population oscillates with damped amplitude. While in overdamping regime, the excited population decays exponentially.
These two regimes can be distinguished by the discriminant of the characteristic equation.

For $D_0<0$ with $\Delta=0$ or $D>0$ with $\Delta\ne0$, the system is in the damped oscillating regime.
On the contrary, for $D_0\ge0$ with $\Delta=0$ or $D\le0$ with $\Delta\ne0$, the system is in the overdamping regime.
After some algebra, the criteria for determining the damped oscillating regime can be summarized as
\begin{equation}\label{Eq_damp_osci_cond}
  \Delta^2\gamma^4+(8\Delta^4-20\Delta^2\Omega^2-\Omega^4)\gamma^2+16(\Delta^2+\Omega^2)^3>0.
\end{equation}
Otherwise, when Eq.~\eqref{Eq_damp_osci_cond} violates, the system is overdamping.
In Fig.~\ref{Fig2}, we give the phase diagram of Rabi oscillating, damped oscillating and overdamping regimes.
The damped oscillating regime is colored in blue and bounded by four curves:\\
(i) the black line,
\begin{equation}
  \gamma=0;
\end{equation}
(ii) the yellow curve $b_2$ ($0\le|\Delta/\Omega|\le\frac{1}{2\sqrt{2}}$),
\begin{equation}
  \gamma=\sqrt{14\Omega^2-4\Delta^2+4\Omega\tfrac{\Omega^2-8\Delta^2}{\Omega+\sqrt{\Omega^2-8\Delta^2}}};
\end{equation}
(iii) and (iv) the yellow curves $b_1$ ($\Delta>0$) and $b_3$ ($\Delta<0$) (both with $0<|\Delta/\Omega|\le\frac{1}{2\sqrt{2}}$),
\begin{equation}
  \gamma=\sqrt{14\Omega^2-4\Delta^2+4\Omega\tfrac{\Omega^2-8\Delta^2}{\Omega-\sqrt{\Omega^2-8\Delta^2}}}.
\end{equation}
The red area corresponds to the overdamping regime.
Here, the intersection point of $b_1$ and $b_2$ is $(\frac{\gamma}{\Omega},\frac{\Delta}{\Omega})=(\frac{9}{\sqrt{6}},\frac{1}{2\sqrt{2}})$ while the intersection point of $b_1$ and $b_3$ is $(\frac{\gamma}{\Omega},\frac{\Delta}{\Omega})=(\frac{9}{\sqrt{6}},-\frac{1}{2\sqrt{2}})$.

The phase diagram is useful for clock frequency estimation.
When the system is overdamping, the population will not oscillate and the clock frequency cannot be inferred.
Therefore, in order to extract the clock frequency via the population oscillation under spontaneous emission, one should make sure that the system is always in the damped oscillating regimes.
The spontaneous emission rate should satisfy the condition that
\begin{equation}\label{gamma}
  0<\gamma<\sqrt{14\Omega^2-4\Delta^2+4\Omega\tfrac{\Omega^2-8\Delta^2}{\Omega+\sqrt{\Omega^2-8\Delta^2}}},
\end{equation}
i.e., $\gamma$ should be on the left side of curve $b_2$ in Fig.~\ref{Fig2}.

Meanwhile, the corresponding analytic solution can be given as (derived from Sec.~\ref{Sec2} B and C)
\begin{equation}\label{Eq_Bloch_damped_oscillating}
  w^Q(t)=C_1e^{\lambda_1 t}+e^{\eta t}\big[C_2\cos(\omega t)+C_3\sin(\omega t)\big]+w^Q_s(t).
\end{equation}
Here, $w^Q_s(t)=-\tfrac{4\Delta^2+\gamma^2}{4\Delta^2+\gamma^2+2\Omega^2}e^{\gamma t}$, $\lambda_1=\frac{1}{3}\gamma+R_1+R_2$, $\eta=\frac{1}{3}\gamma-\frac{1}{2}(R_1+R_2)$ and $\omega=\frac{\sqrt{3}}{2}(R_1-R_2)$ with $R_1=(-\frac{1}{2}q+\sqrt{D})^{1/3}$, $R_2=(-\frac{1}{2}q-\sqrt{D})^{1/3}$, $p=\Delta^2+\Omega^2-\frac{1}{12}\gamma^2$, $q=\frac{1}{108}\gamma(36\Delta^2-18\Omega^2+\gamma^2)$, and $D=\frac{1}{4}q^2+\frac{1}{27}p^3$.
The coefficients $(C_1,C_2,C_3)$ are given by the solution of a linear equation (see Appendix~\ref{SecApp2}),
\begin{equation}
  \left(\!\begin{array}{ccc}
    \lambda_1^2 & \eta^2-\omega^2 & 2\eta\omega \\
    \lambda_1 & \eta & \omega\\
    1 & 1 & 0
  \end{array}\!\right)\!\!\!
  \left(\!\begin{array}{c}
    C_1 \\ C_2 \\ C_3
  \end{array}\!\right)
  =\left(\!\begin{array}{c}
    B_1 \\ B_2 \\ B_3
  \end{array}\!\right),
\end{equation}
where $B_1=-\Delta\Omega u_0+\frac{1}{2}\Omega\gamma v_0-\Omega^2w_0-(w^Q_0+1)\gamma^2$, $B_2=\Omega v_0-(w^Q_0+1)\gamma$, and $B_3=w_0-w^Q_0$ with $w^Q_0=w^Q_s(0)$.

\subsection{Clock frequency estimation in damped oscillating regime\label{Sec3B}}

In this section, we discuss how to estimate the clock frequency of the TLS in the presence of spontaneous emission.
The TLS initials from $\ket{g}$ at time $t=0$, and the population would oscillates when the driving field is applied.
Substituting the initial condition $(u_0,v_0,w_0)^\mathrm{T}=(0,0,-1)^\mathrm{T}$ into Eq.~\eqref{Eq_Pauli_uvw}, we have
\begin{equation}\label{Eq_mean_Pauli}
  \left\{\!\!\begin{array}{ccl}
    \left<\hat\sigma^x\right>&\!\!=\!\!&\cos(\omega_0t)u(t)-\sin(\omega_0t)v(t),\\
    \left<\hat\sigma^y\right>&\!\!=\!\!&\sin(\omega_0t)u(t)+\cos(\omega_0t)v(t),\\
    \left<\hat\sigma^z\right>&\!\!=\!\!&w(t),
  \end{array}\right.
\end{equation}
where $\left<\hat\sigma^z\right>$ represents the population difference between $\ket{e}$ and $\ket{g}$.
When $\gamma$ is modest which satisfies Eq.~\eqref{gamma}, the system is in damped oscillating regime and the evolved excited population $P_e(t)$ reads
\begin{equation}
  P_e(t)=\frac{1+\left<\hat\sigma^z\right>}{2}
  =\frac{1+w(t)}{2}
  =\frac{1+w^Q(t)e^{-\gamma t}}{2}.
\end{equation}
with $w^Q(t)$ being given according to Eq.~\eqref{Eq_Bloch_damped_oscillating}.
Here, the evolution of the excited population $P_e(t)$ is determined by both detuning $\Delta$ and spontaneous emission rate $\gamma$.

If the system is at resonance, i.e., $\omega_D=\omega_0$, the spontaneous emission shrinks the amplitude (or the contrast) of the excited population, see Fig.~\ref{Fig3}~(a).
When $\gamma=0$, the excited population oscillates sinusoidally from 0 to 1 with Rabi frequency $\Omega$.
When $\gamma>0$, the excited population oscillates with damped amplitudes. The excited population would cease to oscillate when the evolution time is long enough.
As $\gamma$ increases, the maximal amplitude decreases rapidly.

The detuning affects both the amplitude (or the contrast) and the period of $P_e(t)$. The evolution of the excited population with different detuning under $\gamma=0.005$ is shown in Fig.~\ref{Fig3}~(b).
As the detuning $\Delta$ gets larger, the maximal amplitude drops and the period of the oscillation becomes shorter.

From the responses of the excited population with different driving frequency $\omega_D$, one can extract the information of clock frequency $\omega_0$.
We calculate the evolved excited population with different detuning $\Delta$ and pick up the maximal amplitude $P_{e,\text{max}}$.
In Fig.~\ref{Fig3}~(c), we show the spectrum of maximal excited population $P_{e,\text{max}}$ versus detuning $\Delta$ with different spontaneous emission rates $\gamma$.
The peaks with different $\gamma$ are all centered at $\Delta=0$, which indicates that no additional frequency shift is induced by the spontaneous emission and $\omega_0$ can be inferred by tuning the driving frequency $\omega_D$ with largest $P_{e,\text{max}}$.

However, the height of the peak decreases as the effects of spontaneous emission becomes stronger.
The peak also becomes less sharper when $\gamma$ increases.
It means that, the process of spontaneous emission would have a negative influence on the estimated precision of $\omega_0$.
To characterize the estimated precision of $\omega_0$ quantitatively, we use the full width at half maximum
(FWHM) of the spectrum denoted by $\Delta\omega_{\text{FWHM}}$.
In the inset of Fig.~\ref{Fig3}~(c), we show the dependence of relative FWHM $\Delta\tilde{\omega}_{\text{FWHM}}$ on the spontaneous emission $\gamma$.
Here, the relative FWHM is defined as the FWHM respect to the one for $\gamma=0$,
\begin{equation}
  \Delta\tilde{\omega}_{\text{FWHM}}(\gamma)=\frac{\Delta\omega_{\text{FWHM}}(\gamma)}{\Delta\omega_{\text{FWHM}}(\gamma=0)}.
\end{equation}
It is shown that, $\Delta\tilde{\omega}_{\text{FWHM}}$ gets larger when $\gamma$ increases, which indicates that the estimated precision of the clock frequency becomes worse when the spontaneous emission gradually comes into play.
Although the spontaneous emission reduces the measurement precision, the clock frequency is not shifted and can still be estimated when the spontaneous emission rate is modest.

\section{Summary\label{Sec4}}

In summary, we have explored the dynamical evolution of a driven TLS under spontaneous emission and illustrate how to perform the clock frequency estimation based on this model.
We derive the Bloch equations for the Lindblad equations in the Dirac picture and give the completely and exactly analytical solutions with closed-form expressions.
In the absence of spontaneous emission, our results recover the perfect Rabi oscillation.
In the presence of spontaneous emission, the system may be in damped oscillating or overdamping regime dependent on both the spontaneous emission rate and the detuning.
We analytically give the boundaries of these regimes and show that how a driven TLS under spontaneous emission can be used for clock frequency estimation.
We find that, the spontaneous emission does not cause additional frequency shift but reduces the estimated precision.
Our results are of potential applications in quantum frequency estimation~\cite{Ivanov2013,Malossi2013,Tian2015,Ivanov2015} and quantum control~\cite{Greilich2009,Poem2011,Bason2012,Sauer2013,Daems2013} of driven TLS's under spontaneous emission.

\acknowledgments

This work is supported by the National Natural Science Foundation of China (Grant No. 11374375, 11574405, 11465008).
Jiahao Huang is partially supported by National Postdoctoral Program for Innovative Talents of China (Grant No. BX201600198).

\appendix

\section{Coefficients for analytical solutions of Lindblad equation under $\Delta=0$\label{SecApp1}}

The coefficients appeared in the solutions to the Lindblad equation are determined by the initial conditions
$\mathbf{R}^Q(t=0)=\mathbf{R}(t=0)=(u_0,v_0,w_0)^\mathrm{T}$ with $u_0,v_0,w_0\in\mathbb{R}$.
We denote $g(t)=w^Q(t)-f_0e^{\gamma t}$ and $f_0=-\tfrac{\gamma^2}{\gamma^2+2\Omega^2}$.
Since
\begin{equation}
  \left\{\!\begin{array}{l}
    u^Q(t)=u_0e^{\frac{1}{2}\gamma t}, \\
    v^Q(t)=\frac{1}{\Omega}[g'(t)+\gamma(f_0+1)e^{\gamma t}], \\
    w^Q(t)=g(t)+f_0e^{\gamma t},
  \end{array}\right.
\end{equation}
we have
\begin{equation}
  \left(\!\begin{array}{c}
    g'(0) \\ g(0)
  \end{array}\!\right)
  =\left(\!\begin{array}{c}
    \Omega v_0-\gamma(f_0+1) \\ w_0-f_0
  \end{array}\!\right).
\end{equation}
Thus the coefficients $(C_1,C_2)$ satisfy with \\
(\romanNum{1}) for $D_0>0$,
\begin{equation}\label{Eq_zero_A1}
  \left(\!\begin{array}{cc}
    \lambda_1 & \lambda_2 \\
    1 & 1
  \end{array}\!\right)
  \left(\!\begin{array}{c}
    C_1 \\ C_2
  \end{array}\!\right)
  =\left(\!\begin{array}{c}
    g'(0) \\ g(0)
  \end{array}\!\right);
\end{equation}
(\romanNum{2}) for $D_0=0$,
\begin{equation}\label{Eq_zero_A2}
  \left(\!\begin{array}{cc}
    \eta & 1 \\
    1 & 0
  \end{array}\!\right)
  \left(\!\begin{array}{c}
    C_1 \\ C_2
  \end{array}\!\right)
  =\left(\!\begin{array}{c}
    g'(0) \\ g(0)
  \end{array}\!\right);
\end{equation}
(\romanNum{3}) for $D_0<0$,
\begin{equation}\label{Eq_zero_A3}
  \left(\!\begin{array}{cc}
    \eta & \omega \\
    1 & 0
  \end{array}\!\right)
  \left(\!\begin{array}{c}
    C_1 \\ C_2
  \end{array}\!\right)
  =\left(\!\begin{array}{c}
    g'(0) \\ g(0)
  \end{array}\!\right).
\end{equation}
Solving the linear equations~\eqref{Eq_zero_A1},~\eqref{Eq_zero_A2},~and~\eqref{Eq_zero_A3} gives the values for the coefficients $(C_1,C_2)$.
Here, we only give results for $D_0<0$ which is concerned in the main text,
\begin{equation}
  \left\{\begin{array}{l}
    C_1=w_0-f_0, \\
    C_2=\frac{1}{\omega}\left[\Omega v_0-\eta(w_0-f_0)-\gamma(f_0+1)\right].
  \end{array}\right.
\end{equation}

\section{Coefficients for analytical solutions of Lindblad equation under $\Delta\ne0$\label{SecApp2}}

Similarly to Appendix~\ref{SecApp1}, we denote $g(t)=w^Q(t)-f_0e^{\gamma t}$,
$f_0=-\tfrac{4\Delta^2+\gamma^2}{4\Delta^2+\gamma^2+2\Omega^2}$, and $f_1=-\tfrac{4\Delta^2\Omega^2}{4\Delta^2+\gamma^2+2\Omega^2}$.
Since
\begin{equation}
  \left\{\!\begin{array}{l}
    u^Q(t)=-\frac{1}{\Delta\Omega}[g''(t)-\frac{1}{2}\gamma g'(t)
    +\Omega^2g(t)+f_1e^{\gamma t}], \\
    v^Q(t)=\frac{1}{\Omega}[g'(t)+\gamma(f_0+1)e^{\gamma t}], \\
    w^Q(t)=g(t)+f_0e^{\gamma t},
  \end{array}\right.\!\!\!\!\!\!\!
\end{equation}
we have
\begin{equation}
  \left(\!\begin{array}{c}
  g''(0) \\ g'(0) \\ g(0)
  \end{array}\!\right)
  \!\!=\!\!\left(\!\begin{array}{c}
  -\Omega(\Delta u_0-\frac{1}{2}\gamma v_0+\Omega w_0)-\gamma^2(f_0+1)
  \\ \Omega v_0-\gamma(f_0+1) \\ w_0-f_0
  \end{array}\!\right).~~~~~~~
\end{equation}
Thus the coefficients $(C_1,C_2,C_3)$ satisfy with \\
(\romanNum{1}) for $D=0$ and $p=0$,
\begin{equation}\label{Eq_nonzero_B1}
  \left(\!\begin{array}{ccc}
    \eta^2 & 2\eta & 2 \\
    \eta & 1 & 0 \\
    1 & 0 & 0
  \end{array}\!\right)
  \left(\!\begin{array}{c} C_1 \\ C_2 \\ C_3 \end{array}\!\right)
  =\left(\!\begin{array}{c} g''(0) \\ g'(0) \\ g(0) \end{array}\!\right);
\end{equation}
(\romanNum{2}) for $D=0$ and $p\ne0$,
\begin{equation}\label{Eq_nonzero_B2}
  \left(\!\begin{array}{ccc}
    \lambda_1^2 & \lambda_2^2 & 2\lambda_2 \\
    \lambda_1 & \lambda_2 & 1 \\
    1 & 1 & 0
  \end{array}\!\right)
  \left(\!\begin{array}{c} C_1 \\ C_2 \\ C_3 \end{array}\!\right)
  =\left(\!\begin{array}{c} g''(0) \\ g'(0) \\ g(0) \end{array}\!\right);
\end{equation}
(\romanNum{3}) for $D<0$,
\begin{equation}\label{Eq_nonzero_B3}
  \left(\!\begin{array}{ccc}
    \lambda_1^2 & \lambda_2^2 & \lambda_3^2 \\
    \lambda_1 & \lambda_2 & \lambda_3 \\
    1 & 1 & 1
  \end{array}\!\right)
  \left(\!\begin{array}{c} C_1 \\ C_2 \\ C_3 \end{array}\!\right)
  =\left(\!\begin{array}{c} g''(0) \\ g'(0) \\ g(0) \end{array}\!\right);
\end{equation}
(\romanNum{4}) for $D>0$,
\begin{equation}\label{Eq_nonzero_B4}
  \left(\!\begin{array}{ccc}
    \lambda_1^2 & \eta^2-\omega^2 & 2\eta\omega \\
    \lambda_1 & \eta & \omega \\
    1 & 1 & 0
  \end{array}\!\right)
  \left(\!\begin{array}{c} C_1 \\ C_2 \\ C_3 \end{array}\!\right)
  =\left(\!\begin{array}{c} g''(0) \\ g'(0) \\ g(0) \end{array}\!\right).
\end{equation}
Solving the linear equations~\eqref{Eq_nonzero_B1},~\eqref{Eq_nonzero_B2},~\eqref{Eq_nonzero_B3},~and~\eqref{Eq_nonzero_B4} gives the values for the coefficients $(C_1,C_2,C_3)$.
Here, we only give results for $D>0$ which is concerned in the main text,
\begin{equation}\label{Eq_nonzero_C1C2C3}
  \left\{\begin{array}{l}
    C_1=-\frac{\Omega\Delta}{(\lambda_1-\eta)^2+\omega^2}u_0
    +\frac{\Omega(\frac{1}{2}\gamma-2\eta)}{(\lambda_1-\eta)^2+\omega^2}v_0
    +\frac{\eta^2+\omega^2-\Omega^2}{(\lambda_1-\eta)^2+\omega^2}w_0 \\
    \quad\quad+\frac{\gamma(2\eta-\gamma)(f_0+1)-f_0(\eta^2+\omega^2)}{(\lambda_1-\eta)^2+\omega^2}, \\
    C_2=w_0-f_0-C_1, \\
    C_3=\frac{1}{\omega}\left[\Omega v_0-\eta(w_0-f_0)-\gamma(f_0+1)
    +(\eta-\lambda_1)C_1\right].
  \end{array}\right.
\end{equation}

The results in Sec.~\ref{Sec2C} and Appendix~\ref{SecApp2} are for $\Delta\ne0$.
But the calculations should be also applicable for $\Delta=0$.
To reveal this point, we assume $\Delta=0$ in the following discussion.
Since $\Delta=0$, we have
$q=\frac{1}{108}\gamma(\gamma^2-18\Omega^2)$,
$D=-\frac{1}{108}\Omega^4 D_0$ ($D_0=\frac{1}{4}\gamma^2-4\Omega^2$),
and $\omega=\frac{\sqrt{3}}{2}\big[(-\frac{q}{2}+\sqrt{D})^{1/3}+(\frac{q}{2}+\sqrt{D})^{1/3}\big]$.
Introducing $x=\frac{1}{2\Omega}\sqrt{-D_0}$ (where $0<x\le 1$),
we have $D_0=-4\Omega^2 x^2$, $D=\frac{1}{27}\Omega^6x^2$,
$\gamma=4\Omega\sqrt{1-x^2}$,
$q=-\frac{2}{27}\Omega^3(1+8x^2)\sqrt{1-x^2}$,
and
\begin{equation}
  \omega=\tfrac{1}{2}\Omega\left(S_1+S_2\right).
\end{equation}
Here, we introduce
\begin{equation}
  \left\{\!\!\begin{array}{l}
    S_1=\left[x+\tfrac{\sqrt{3}}{9}(1+8x^2)\sqrt{1-x^2}\right]^{1/3}, \\
    S_2=\left[x-\tfrac{\sqrt{3}}{9}(1+8x^2)\sqrt{1-x^2}\right]^{1/3}.
  \end{array}\right.
\end{equation}
Then $S_1+S_2=\frac{2\omega}{\Omega}$, $S_1S_2=\frac{1}{3}(4x^2-1)$.
By $(S_1+S_2)^3=S_1^3+S_2^3+3S_1S_2(S_1+S_2)$,
we have $(\frac{2\omega}{\Omega})^3=2x+(4x^2-1)\frac{2\omega}{\Omega}$,
i.e. $(\omega-\Omega x)(4\omega^2+4\omega\Omega x+\Omega^2)=0$.
Therefore,
\begin{equation}
  \omega=\Omega x=\tfrac{1}{2}\sqrt{-D_0}.
\end{equation}
Similarly, $\eta=\frac{1}{3}\gamma-\frac{1}{2}(R_1+R_2)=\frac{1}{4}\gamma$.
Since $\Delta=0$, $\frac{1}{2}\gamma-2\eta=0$, $\eta^2+\omega^2-\Omega^2=0$,
and $\gamma(2\eta-\gamma)(f_0+1)-f_0(\eta^2+\omega^2)=0$, we have
\begin{equation}
  \left\{\begin{array}{l}
    C_1=0, \\
    C_2=w_0-f_0, \\
    C_3=\frac{1}{\omega}\left[\Omega v_0-\eta(w_0-f_0)-\gamma(f_0+1)\right]
  \end{array}\right.
\end{equation}
[from Eq.~\eqref{Eq_nonzero_C1C2C3}].
All these recover the results in Sec.~\ref{Sec2B} and Appendix~\ref{SecApp1}.

\end{document}